\documentclass[sn-mathphys,Numbered]{sn-jnl}
\usepackage{colortbl}
\usepackage{xcolor}
\usepackage{amsmath}
\usepackage{graphicx}
 
\usepackage{hyperref}
\usepackage{a4,fullpage}
\usepackage{threeparttable}  
\usepackage{multicol}
\usepackage{multirow}
\usepackage{tabularx}
\usepackage{graphicx} 
\usepackage{multirow}  
\usepackage{amsmath,amssymb,amsfonts} 
\usepackage{amsthm} 
\usepackage{mathrsfs} 
\usepackage[title]{appendix} 
\usepackage{textcomp} 
\usepackage{manyfoot} 
\usepackage{booktabs} 
\usepackage{algorithm} 
\usepackage{algorithmicx} 
\usepackage{algpseudocode} 
\usepackage{listings}

\begin{document}

\title[Leveraging Artificial Intelligence Technology for Mapping Research to Sustainable Development Goals: A Case Study]{Leveraging Artificial Intelligence Technology for Mapping Research to Sustainable Development Goals: A Case Study}

\author*[1]{\fnm{Hui} \sur{Yin}}\email{huiyin@swin.edu.au}

\author[1]{\fnm{Amir} \sur{Aryani}}\email{aaryani@swin.edu.au}

\author[1]{\fnm{Gavin} \sur{Lambert}}\email{glambert@swin.edu.au}

\author[1]{\fnm{Marcus} \sur{White}}\email{marcuswhite@swin.edu.au}

\author[2]{\fnm{Luis} \sur{Salvador-Carulla}}\email{luis.salvador-carulla@canberra.edu.au}

\author[3]{\fnm{Shazia} \sur{Sadiq}}\email{shazia@itee.uq.edu.au}
\author[4]{\fnm{Elvira} \sur{Sojli}}\email{e.sojli@unsw.edu.au}
\author[5]{\fnm{Jennifer} \sur{Boddy}}\email{j.boddy@griffith.edu.au}
\author[1]{\fnm{Greg} \sur{Murray}}\email{gwm@swin.edu.au}
\author[4]{\fnm{Wing} \sur{Wah Tham}}\email{w.tham@unsw.edu.au}

\affil[1]{Swinburne University of Technology, Australia}
\affil[2]{University of Canberra, Australia}
\affil[3]{University of Queensland, Australia}
\affil[4]{University of New South Wales, Australia}
\affil[5]{Griffith University, Australia}


\abstract{
The number of publications related to the Sustainable Development Goals (SDGs) continues to grow. These publications cover a diverse spectrum of research, from humanities and social sciences to engineering and health. 
Given the imperative of funding bodies to monitor outcomes and impacts, linking publications to relevant SDGs is critical but remains time-consuming and difficult given the breadth and complexity of the SDGs.
A publication may relate to several goals (interconnection feature of goals), and therefore require multidisciplinary knowledge to tag accurately. Machine learning approaches are promising and have proven particularly valuable for tasks such as manual data labeling and text classification. 
In this study, we employed over 82,000 publications from an Australian university as a case study. We utilized a similarity measure to map these publications onto Sustainable Development Goals (SDGs). 
Additionally, we leveraged the OpenAI GPT model to conduct the same task, facilitating a comparative analysis between the two approaches. Experimental results show that about 82.89\% of the results obtained by the similarity measure overlap (at least one tag) with the outputs of the GPT model. The adopted model (similarity measure) can complement GPT model for SDG classification. Furthermore, deep learning methods, which include the similarity measure used here, are more accessible and trusted for dealing with sensitive data without the use of commercial AI services or the deployment of expensive computing resources to operate large language models. Our study demonstrates how a crafted combination of the two methods can achieve reliable results for mapping research to the SDGs.}




\maketitle
\section{Introduction}
In 2015, all Member States of the United Nations adopted the 2030 Agenda for Sustainable Development, which provided a blueprint for peace and prosperity for people and the planet. 
All developed and developing countries were called upon to act in a global partnership to achieve the 17 Sustainable Development Goals (SDGs). The global indicator framework for Sustainable Development Goals includes 17 goals, 169 more detailed SDG targets, and 231 unique indicators.\footnote{A description and list of goals is available at \url{https://sdgs.un.org/goals}. More details on the SDG goals, targets, and indicators are available at \url{https://unstats.un.org/sdgs/indicators/indicators-list/}.}
The goals foster global cooperation and provide a framework for sustainable economic growth, social inclusion, and environmental protection.
Academia, government, and industry all play a crucial role in promoting the realization of these goals. 
In academia, experts review the extant literature, develop research programs, analyze and publish their findings. 
These results can inform and contribute to the achievement of the 17 goals.
Researchers benefit from SDG-related publications, which are large and growing daily, to understand the current state of research, plan future projects and identify potential collaboration with policy makers and other stakeholders.
Publications can provide information for reporting, communicating, and showcasing university contributions to the SDGs. 
Furthermore, publications serve as essential indicators for monitoring the progress of each country, region, and institution in achieving the SDGs~\cite{kestin2017getting}. 
When the literature is viewed critically, it helps to paint a comprehensive picture of progress towards the SDGs~\cite{kestin2017getting}.
However, it is difficult for public and private research institutions to identify their past and current research that relates to the SDGs~\cite{mori2019implementing,nakamura2019navigating} because 
categorizing publications into the corresponding 17 goals is challenging and requires multidisciplinary knowledge. 
For example, the concept of ``Green Infrastructure" as applied within the European Union (EU) can result in diverse interpretations and viewpoints across various disciplines.\footnote{A comprehensive description of the green infrastructure concept is available here \url{https://environment.ec.europa.eu/topics/nature-and-biodiversity/green-infrastructure_en}.} Experts in fields such as environmental science, urban planning, and economics may possess distinct understandings of the scope and alignment of ``Green Infrastructure" with broader sustainable development objectives. Indeed, as sustainability research is often cross-disciplinary, a single publication can belong to multiple goals and address multiple research fields.
Considering the burgeoning volume of publications and the difficulty in accurate and seamless classification, it is not practical to complete the mapping task manually by an expert.
There are also limitations when using traditional keyword search techniques~\cite{ElsevierKeywords,korfgen2018sa}.
In the first place, query methods are based on surveys, so they are subjective in nature and there is an ongoing dispute over the accuracy of such queries~\cite{armitage2020mapping}. Second, it is impossible to reflect the degree of correspondence between publications and SDGs through the query method~\cite{zhang2020matching}. Third, using the query method, the obtained publications relate to the current field; nevertheless, their relevance to the SDGs cannot be guaranteed. While they share certain keywords with the SDG goals, these publications may not necessarily align with the intended SDG context.
\begin{figure}[h]
\centering
\includegraphics[width=0.8\linewidth]{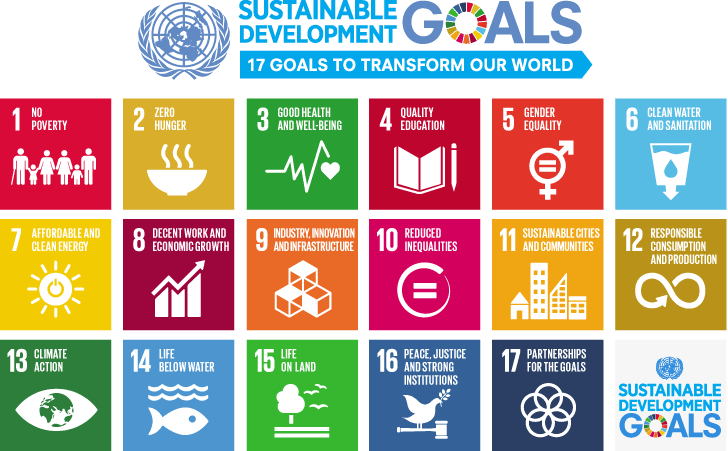}
\caption{\label{fig:SDG}Sustainable development goals from United Nations Agenda 2030.}
\end{figure}

In recent years, machine learning has proven particularly valuable for tasks in natural language processing (NLP).
When dealing with large volumes of publications an automatic classification method is desirable to efficiently and accurately facilitate the assigning of articles to the corresponding goals.
By establishing and using such tools, research institutions would be able to determine and align their research strengths and gaps towards the SDGs. Considering the use of machine learning technology, this task is a multi-label classification task in NLP.\footnote{An extended description of multi-label classifications is available from \url{https://en.wikipedia.org\/wiki\/Multi-label\_classification}.} Multi-label classification of data remains a challenging problem. Because of the complexity of the data, it is sometimes difficult to infer information about classes that are not mutually exclusive.

In this work, using publications derived from Swinburne University of Technology staff as a case study, we designed an automatic method to map publications to the SDGs using Artificial intelligence techniques.
Swinburne University of Technology\footnote{https://www.swinburne.edu.au/} is an Australian university in Melbourne, Victoria. Established in 1908, it is a dual-sector institution providing higher education and vocational training and is recognized for its strong commitment to innovation, industry engagement, and applied research.
We summarize our contributions as follows: First, we harnessed AI-based similarity measures to effectively map publications onto the Sustainable Development Goals (SDGs). Subsequently, we explored the possibility of employing the Generative Pre-trained Transformer (GPT) for this task. We extensively delved into the emerging challenges and the innovative solutions we employed to overcome them. Lastly, we conducted experiments to demonstrate that our approach is user-friendly, readily implementable, and characterized by efficient computational demands. Furthermore, our method yields reliable outputs and demonstrates performance on par with the GPT model.

\section{Related work}
Recently, there has been a notable growing trend of employing AI techniques for SDG-related research and activities. 
Mainstream AI approaches fall into two types: supervised learning and unsupervised learning.
Supervised learning can only be applied  with a big-enough domain-specific training set.
Arash et al.~\cite{DBLP:journals/scientometrics/HajikhaniS22} trained a machine learning model (Naive Bayes, Linear Support Vector Machine, Logistic Regression) on SDGs classification of science publications, then used this classifier to classify the SDGs relevancy in patent documents.
To construct the training set for the classification algorithm, they applied a keyword search to map scientific publications from 2015 to 2020 from the SCOPUS publication database. The extracted list of keywords was consistent with existing taxonomies \cite{jia2019visualizing,vatananan2019bibliometric,ElsevierKeywords}.
For the text modeling, they adopted TF-IDF, Word2vec, and Doc2vec. 
The highest overall accuracy (f-score) was achieved by "Word2vec and logistic regression" models. 

Guisiano and Chiky~\cite{guisiano2021automatic} linked project descriptions to one or more SDGs.
For the training dataset, Guisiano and Chiky extracted 169 targets from each of the 17 SDGs and then applied data enrichment technology to increase the SDG target size from 169 to 6,017. 
A BERT (Bidirectional Encoder Representations from Transformers) classifier was trained on the dataset for multi-label text classification. 
They tested the model's performance results by texts from the SDG pathfinder.\footnote{\url{https://sdg-pathfinder.org}.}
Matsui et al.~\cite{matsui2022natural} also uses the BERT classifier, employing a pre-trained Japanese BERT model.
In this work, Matsui et al. built a classifier capable of semantic mapping practices and issues in SDGs content and visualized the co-occurrence of goals. They designed a data framework to extract data from documents published by official organizations and multi-labels corresponding to the SDGs.
Pre-trained Japanese BERT models were fine-tuned on multi-label text classification tasks, and nested cross-validation was performed to optimize the hyperparameters.

The alternative methodology is to employ unsupervised learning. Zhang et al.~\cite{DBLP:conf/dsaa/ZhangVSZ20} were the first to use a deep learning approach for such a task in 2020.
They proposed a hierarchical structure to match research publications to SDGs using the three levels of the SDGs: Goals, Targets, and Indicators, respectively.
Their real-world data included 618,675 publications with labeled SDGs (Goal only) through the set of queries developed by Elsevier~\cite{ElsevierKeywords}. 
The model predicts the possibility of SDG indicators, targets, and goals for each publication, making it suitable for mapping tasks that require a fine-grained division of SDGs.
% The study of Toetzke et al.~\cite{toetzke2022monitoring} clustered ~3.2 million descriptions of aid events conducted between 2000 and 2019 to monitor new topics and identify unexplored spatiotemporal disparities. They employed a paragraph vector model to embed descriptions into document vectors, followed by clustering using the K-Means algorithm, and found 173 activity clusters.
Another approach has used Latent Dirichlet Allocation (LDA) technology to cluster 267 individual Department of Economic and Social Affairs (DESA) publications of United Nations Secretariat. The number of topics was set to 18, 17 of which are SDGs and the last one captures the commonality~\cite{lafleur2019art}.
Morales-Hernández et al.~\cite{morales2022comparison} conducted a study comparing the performance of a text classification model with Label Power sets (LP) and Support Vector Machines (SVMs) with DistilBERT on five imbalanced and balanced datasets. 
Performance metrics have confirmed that LP-SVM remains an effective tool for classifying multi-label texts with remarkable results.
Mishra et al.~\cite{mishra2023bibliometric} recently conducted a bibliometric analysis of publications related to the Sustainable Development Goals (SDGs), utilizing the Web of Science (WoS) core collection citation database as the primary source. They employed advanced search keywords such as ``Sustainable Development Goals," ``SDGs," and ``Millennium Development Goals" and retrieved 12,176 publications spanning from 2015 to 2022.
They conducted a comprehensive bibliometric analysis encompassing literature trend analysis, characteristics of research publications, the most popular and productive journals, highly productive countries, and collaborative efforts among authors.
Their research scrutinized the status and trends within SDG-related literature from a macro level. However, it is worth noting that some bibliometric data may have been missed if this publication did not include the same keywords used in the query~\cite{ElsevierKeywords,korfgen2018sa}, as we discussed in the introduction section.

\section{Insights of the publications}
\begin{figure}[h]
\centering
\includegraphics[width=0.8\linewidth]{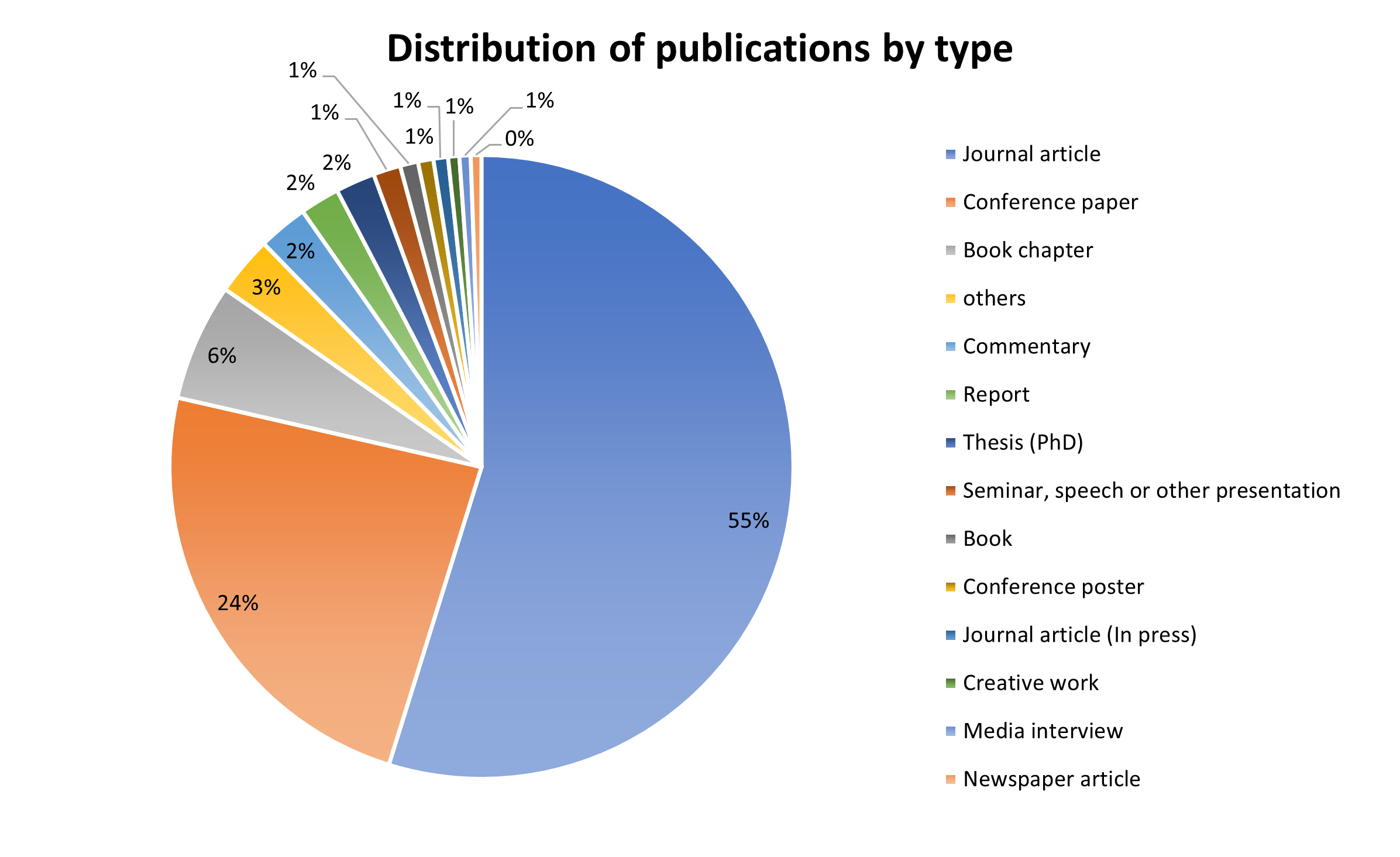}
\caption{\label{fig:PublicationDistribution}Distribution of publications by type in the case study.}
\end{figure}

In this case study, we utilized over 82,000 publications from the Swinburne University of Technology research data bank, spanning the period from 1967 to 2023. 
This sample included many publication types, and figure~\ref{fig:PublicationDistribution} shows the share of all publication types.
Among these, journal articles account for the largest proportion, 55\%, followed by conference papers, accounting for 24\%. A smaller fraction comes from ``other" publication types, such as curated exhibitions or events, creative work, datasets, and less frequent, categories.
Not all publications were SDG-related, so we filtered for unrelated publications before further analysis.
The following section describes the cosine similarity measure and how we use it to remove publications that are not relevant to the SDGs.

\subsection{AI-based similarity measure}
Machine learning models operate on numerical data, so converting words to embeddings is crucial in NLP.
Word embeddings are continuous-valued vector representations of words that are learned through unsupervised training on large textual datasets. These embeddings capture contextual information and include positional information within a high-dimensional space. 
During the training process, words that appear in similar contexts are mapped to nearby points in the high-dimensional space.
% In a high-dimensional embedding space, the cosine similarity between two vectors measures the cosine of the angle between them.
In mathematics, cosine similarity is the cosine value of the angle between two vectors projected in a multidimensional space, which can be used to measure the similarity of documents in NLP applications. The score ranges from -1 to 1; the smaller the angle, the higher the similarity.
Cosine similarity is widely used in various NLP tasks, such as information retrieval, document similarity analysis, and clustering. It helps to quantify the similarity between words or documents in the vector space, allowing NLP models to leverage this similarity information for various language processing tasks.
For example, we might see that the following word embeddings equations are valid: ``Clean Energy - Fossil Fuels + Renewable Sources = Sustainable Power", ``Decent Work - Exploitative Labor + Fair Employment Practices = Economic Growth."

The formula for calculating cosine similarity is as follows:
\begin{equation}
\label{equ:cosine}
 \cos (\theta ) =   \dfrac {A \cdot B} {\left\| A\right\| _{2}\left\| B\right\| _{2}},      
\end{equation}

\noindent where:
$A$ and $B$ are the vector representations of the two sentences.
$A \cdot B$ represents the dot product of vectors A and B.
$||A||$ and $||B||$ represent the Euclidean norms (or magnitudes) of vectors A and B, respectively.

\begin{figure}[h]
\centering
\includegraphics[width=1\linewidth]{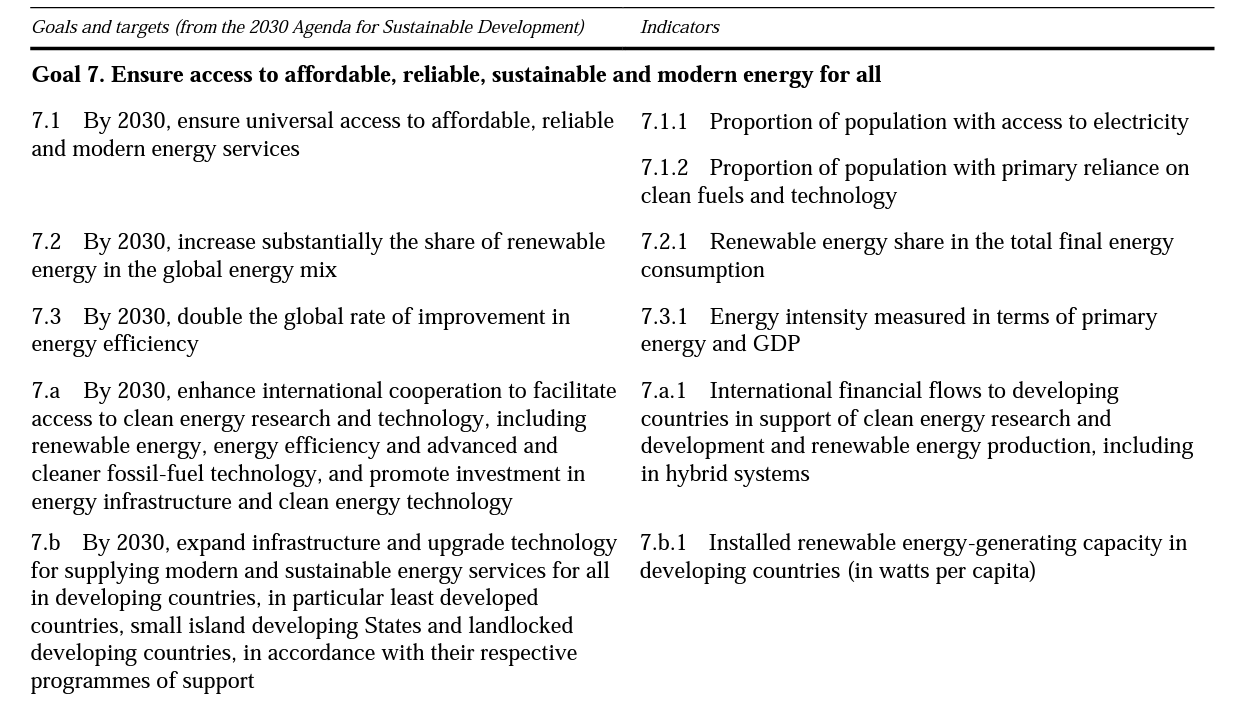}
\caption{\label{fig:Goal7} Example of Goal 7 from the sustainable development goals from United Nations Agenda 2030, each of which includes Goal, Target and Indicator.}
\end{figure}

We designed the following method: Applying a language model to generate publication vectors and SDG vectors, we calculate the cosine similarity between a publication vector and the 17 target vectors individually. This process allows us to obtain the cosine similarity score between a publication and the 17 Sustainable Development Goals, reflecting the similarity of the publication's content to these goals. 
For the publications, given the length of research articles, it is impractical, resource-intensive and time-consuming to input the entire article into the language model to generate vectors for categorizing publications. In a publication, the title and abstract are highly condensed and succinctly summarize a publication, and this information is sufficient to represent the content of one publication. Therefore, we combined the title and abstract and input them into the language model to generate the vector for the publication.
The global indicator framework for Sustainable Development Goals includes 17 goals, 169 more detailed SDG targets, and 231 unique indicators.\footnote{More details on the SDG goals, targets, and indicators are available here: \url{https://unstats.un.org/sdgs/indicators/indicators-list/}.}
For the goals, we use the combination of goal, target, and indicator to generate vectors to extract more features from each goal.
Figure~\ref{fig:Goal7} shows an example of the sustainable development goals, targets and indicators for Goal 7 from the United Nations Agenda 2030.

\begin{figure}[h]
\centering
\includegraphics[width=0.6\linewidth]{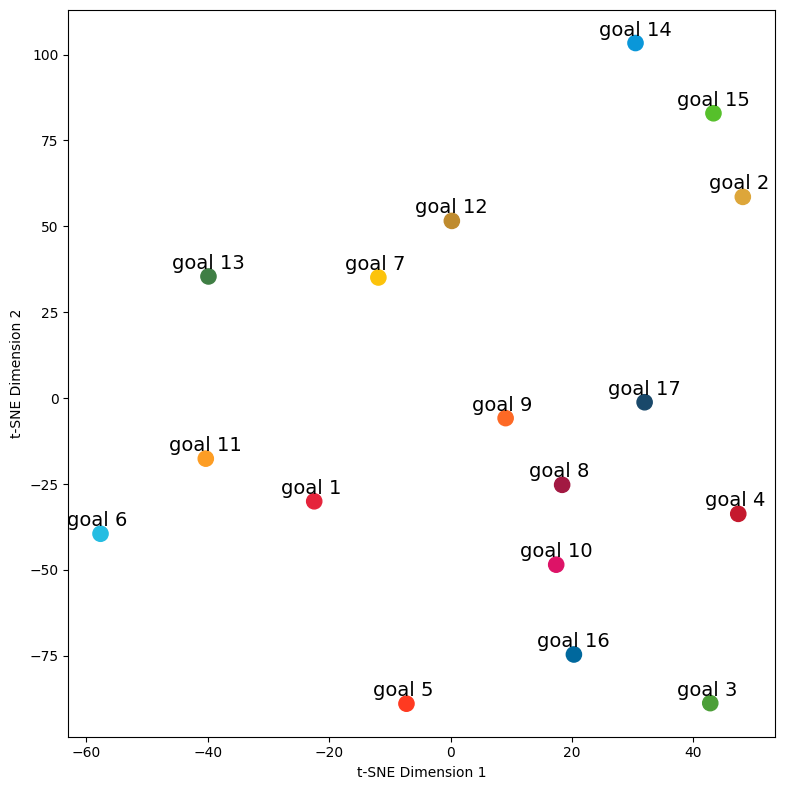}
\caption{\label{fig:GoalPosition}
% Map of the relative position of SDG goals in high-dimensional space transformed to 2D space using tsne.
High-dimensional space visualization of 17 SDG goals using t-SNE dimension reduction.}
\end{figure}

We employed a language model Sentence-BERT\footnote{\url{https://www.sbert.net/\#:~:text=SentenceTransformers\%20is\%20a\%20Python\%20framework,for\%20more\%20than\% 20100\%20languages.}} to generate the 17 goals vectors (in a 768-dimensional dense vector space).
Sentence-BERT (Sentence Embeddings using Siamese BERT Network) is a variant of the original BERT (Bidirectional Encoder Representations from Transformers) model that is specifically designed for generating sentence embeddings. BERT, developed by Google in 2018, is a powerful language representation model that can understand the context of words in a sentence by training on large amounts of text data~\cite{devlin2019bert}.
Consider that the SDGs are deeply interconnected - a lack of progress on one goal hinders progress in others, so we performed dimensionality reduction on the goal vectors to explore the relationship between goals, using t-Distributed Stochastic Neighbor Embedding (t-SNE).
t-SNE enables the visualization of high-dimensional data in two or three dimensions, providing a way to visualize relationships that may be challenging to observe in the original high-dimensional space.
Figure~\ref{fig:GoalPosition} shows the relative positions of 17 goals in 2D dimensions. 
The figure shows that many goals are clearly distinguishable from each other (e.g., the boundaries between goals are clear), but it can be noticed that goals 2 (Zero hunger), goal 14 (Life below water), and goal 15 (Life on land) are relatively close, and their contents are indeed related.
Goal 3 (Good health and well-being), goal 6 (Clean water and sanitation), and goal 14 (Life below water) have the farthest spatial distance because they do not correlate with content; thus the vector representations have the most significant difference.

% For the publications,  we combined the title and abstract to represent a publication.
% We still use the language model Sentence-BERT to generate vectors (embeddings) of publications, ensuring that the obtained vector dimensions are the same as the 17 goals.
After getting the vector of publications and 17 goals, we calculated the cosine angles of each publication's vector to the 17 goals according to equation~\ref{equ:cosine}to calculate the similarity score between the publication and the 17 goals. 
Table~\ref{tab_ExampleCosine} shows some examples of the method outputs.

\begin{table*}[h]
\centering
\caption{The output of the cosine similarity measure.}
\label{tab_ExampleCosine}

\begin{tabular}{m{3cm}|m{7.5cm}|m{4.2cm}}
\hline
\hline
 Authors & Publication & Outputs\\ \hline

Suni Mydock III et al. & Title: Influence of made with renewable energy appeal on consumer behaviour (2017) Abstract: The purpose of this paper is to explore the extent to which consumer purchasing behaviour is influenced by advertised information that a product is made with renewable energy .... ... Originality/value This research is the first of its kind to be conducted in an Australian context, providing findings that assist both firms' and policy-makers' decision-making. & \textbf{Goal 12: 0.44}, \textbf{Goal 7: 0.41}, Goal 13: 0.30, Goal 14: 0.25, Goal 17: 0.23, Goal 9: 0.22, Goal 2: 0.21, Goal 15: 0.2, Goal 3: 0.19, Goal 6: 0.18, Goal 8: 0.17, Goal 5: 0.16, Goal 11: 0.15, Goal 4: 0.15, Goal 10: 0.14, Goal 1: 0.12, Goal 16: 0.12 \\ \hline
Newton, Peter W. &Title: No sustainable population without sustainable consumption (2011) Abstract: In recent weeks, two major federal government strategy papers have been released: 'Our Cities, Our Future: A National Urban Policy for a Productive, Sustainable and Liveable Future' and 'Sustainable Australia---Sustainable Communities: A Sustainable Population Strategy for Australia'.  ... ...  In a growing and urbanising 21st century world it is an unsustainable and inequitable nexus.  & \textbf{Goal 11: 0.60}; \textbf{Goal 12: 0.55}; \textbf{Goal 8: 0.47}; \textbf{Goal 4: 0.46}; \textbf{Goal 6: 0.45}; \textbf{Goal 15: 0.45}; \textbf{Goal 13: 0.44}; \textbf{Goal 1: 0.44}; \textbf{Goal 3: 0.42}; \textbf{Goal 9: 0.41}; \textbf{Goal 14: 0.40}; \textbf{Goal 7: 0.40};  Goal 10: 0.39 ; Goal 17: 0.37 ;  Goal 2: 0.36 ; Goal 16: 0.36; Goal 5: 0.30 \\ 

\hline
\hline
\end{tabular}
\begin{tablenotes}
      \small
      \item Note: The goal in bold is the goal whose confidence level is higher than the threshold, which is the final output of the similarity measure method.
    \end{tablenotes}
\end{table*}

\subsection{Remove irrelevant publications}
\label{sec:RemoveIrrelevant}

\begin{figure}[h]
\centering
\includegraphics[width=0.6\linewidth]{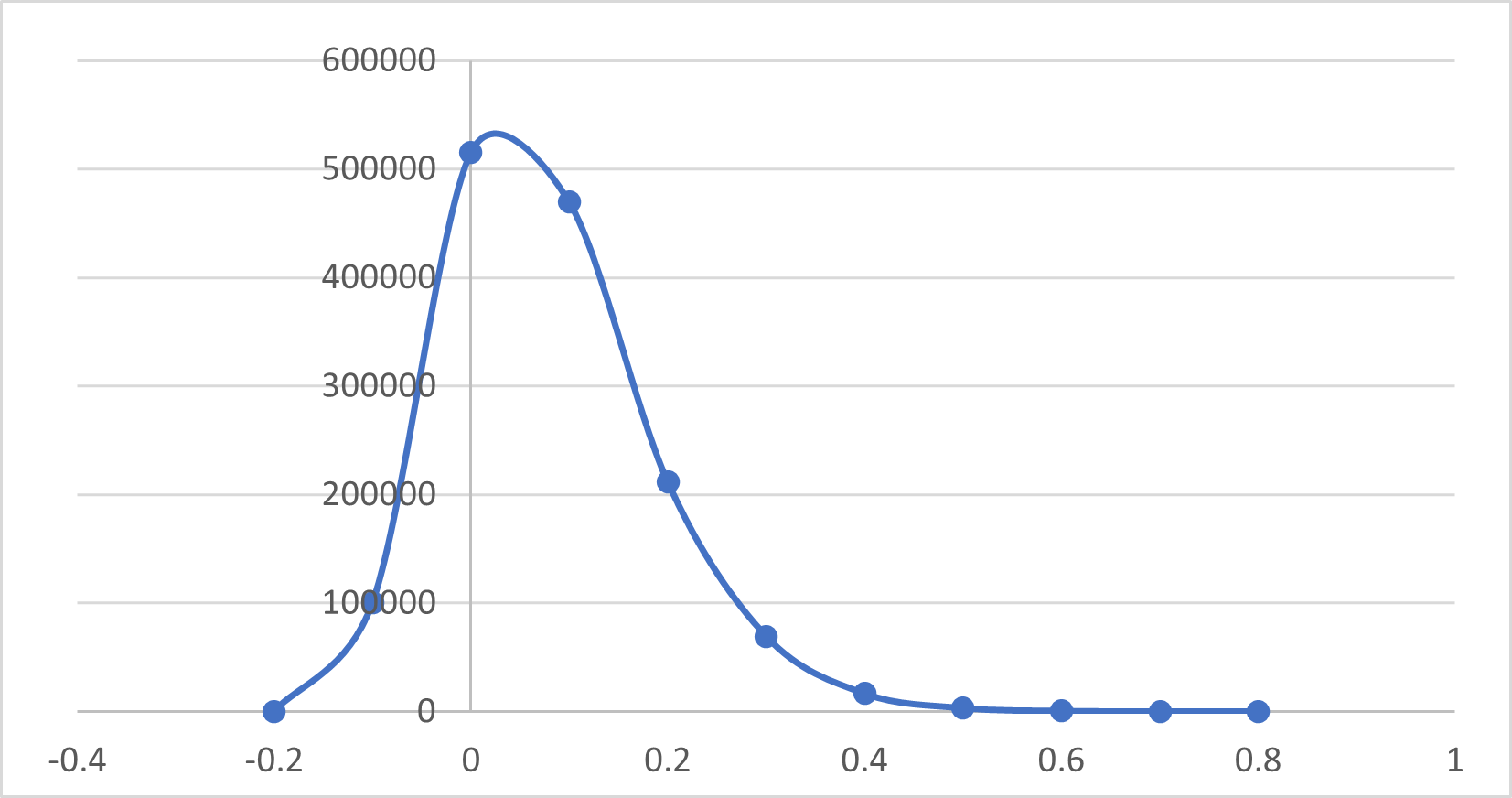}
\caption{\label{fig:Threshold}Distribution of cosine similarity values for all publications using the similarity measure.}
\end{figure}

In the experiment involving 82,649 publications, a similarity measure was employed to calculate the similarity score between each publication and the 17 SDGs. Through this process, we analyzed the distribution of these scores, as illustrated in figure~\ref{fig:Threshold}, with the goal of establishing an appropriate threshold. This threshold enables us to retain only the SDGs for each publication whose similarity scores surpass the predefined threshold.

\begin{figure}[h]
\centering
\includegraphics[width=0.8\linewidth]{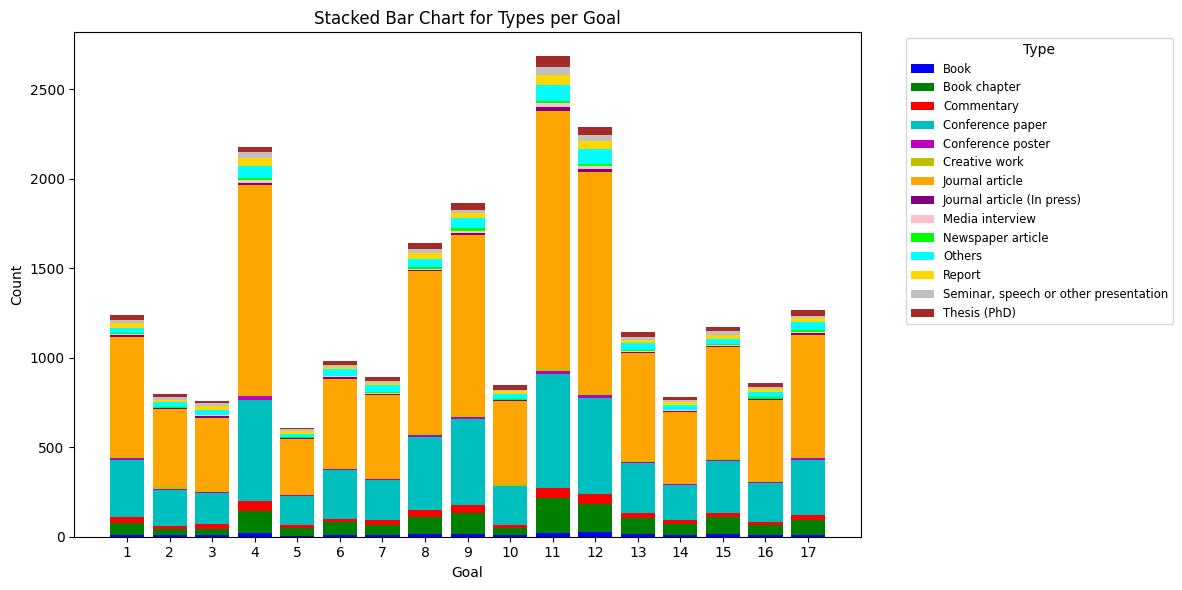}
\caption{\label{fig:BERT_barchart}Publication distribution across 17 SDGs: stacked bar chart of different publication types using similarity measure method.}
\end{figure}

According to the distribution of scores, experts were also invited to conduct random inspection of the results, with 0.4 determined as the threshold.
We found that 7,403 publications were related to the SDGs, accounting for 0.09\% of the collected publications.
Each publication has at least one tag with a similarity score greater than 0.4 and a maximum of 16 tags.
Similarity measures tagged 18,924 SDGs for 7,403 publications. 
We counted the number of publications under each goal, shown in figure~\ref{fig:BERT_barchart}, and we can conclude that the researchers of Swinburne University of Technology conducted research across all SDGs, with the greatest number of publications associated with the following goals: 
\begin{itemize}
    \item Goal 11: Sustainable Cities and Communities
    \item Goal 12: Responsible Consumption and Production
    \item Goal 4: Quality Education
\end{itemize}

\section{GPT for mapping task}
Large language model (LLM) refers to a Pre-trained Language Model (PLM) of significant size, e.g., containing tens or hundreds of billions of parameters, trained on massive text data~\cite{shanahan2023talking}, such as GPT-3~\cite{brown2020language}, BERT~\cite{devlin2019bert}, XLNet~\cite{yang2019xlnet}, T5~\cite{10.5555/3455716.3455856}.
GPT, developed by OpenAI, is known as a game changer for natural language processing (NLP). It can be a valuable tool for researchers in several ways, such as information retrieval, literature review, proofreading and editing, coding and debugging, and idea generation~\cite{zhao2023survey,dong2023survey}.
In addition, GPT also offers a user-friendly interface called ChatGPT, which exhibits remarkable conversational abilities with humans. It can directly provide answers to users' questions, unlike information retrieval search engines like Google and Bing, which require users to filter through result URLs to find the appropriate information.  

The ChatGPT web app (available in your browser) is explicitly designed for chatbot applications. 
It relies on GPT to produce text, such as explaining code, writing poems or answering questions on an extensive and broad range of topics. 
The chat models (GPT model) take a list of messages as input and return a model-generated message as output.
We took advantage of this ability and used the OpenAI API to tag Swinburne University publications to SDGs.

\subsection{Chat completions API}
An example API call looks as follows:

\begin{minipage}[c]{0.95\textwidth}
\lstset{texcl=true,basicstyle=\small\sf,commentstyle=\small\rm,mathescape=true,escapeinside={(*}{*)}}
\begin{lstlisting}
import openai
API_KEY = `your API_KEY'
openai.api_key= API_KEY
openai.ChatCompletion.create(
model="gpt-3.5-turbo",
messages=[
    {"role": ``user", "content": "Analyze the 
    publication and determine the SDGs it 
    aligns with..."}
   ], max_tokens=600
)
\end{lstlisting}
\end{minipage}

There are three necessary elements to use the API, the API\_KEY, model, and prompt.
API\_KEY can be requested on OpenAI.com. 
With the API key, developers can start integrating the GPT API into applications.
In terms of model selection, GPT has a variety of models that can be selected for different purposes.
In our study, we selected the ``gpt-3.5-turbo" version, which aligns with the version underlying ChatGPT's architecture.
The inputs of GPT are referred to as ``prompts".
Our purpose was to use GPT to tag the publications while providing a confidence level and explanation for each specified result to assist in judging whether it is meaningful.
Considering that GPT is charged by token (input and output), the prompt must be concise and clear. Controlling the number of output tokens by setting the max\_tokens parameter can also help control token numbers.
Our final prompt was designed to Analyze the publication and determine the SDGs that it aligned with, provide the confidence levels(\%) for each assigned goal and the reason for assignment. Results listed the goals in descending order of confidence level, with the highest confidence goal listed first. Title:***. Abstract: ***.
According to our final statistics, this prompt can ensure that more than 90\% of the outputs meet the expectations.
\subsection{Errors and solution}
Programs servicing paying users are often interrupted because of overload, where demand is beyond the server's capacity.
Table~\ref{tab_Errors} lists the error messages we encountered during the operation.
\begin{table*}[h]
\centering
\caption{Error messages when running GPT API.}
\label{tab_Errors}

\begin{tabular}{m{1.5cm}|m{13.2cm}}
\hline
\hline
 Formula & Explanation \\ \hline

 Error 1 & The server is overloaded or not ready yet. \\ \hline
  Error 2 & That model is currently overloaded with other requests. You can retry your request, or contact us through our help center at help.openai.com if the error persists. \\ \hline
  Error 3 & Bad gateway. \{``error":\{``code":502,``message":``Bad gateway.",``param":null,``type":``cf\_bad\_gateway"\}\} 502 \{'error': \{'code': 502, `message': `Bad gateway.', `param': None, 'type': 'cf\_bad\_gateway'\}\} \{'Date': 'Tue, 04 Jul 2023 05:23:06 GMT', 'Content-Type': 'application/json', 'Content-Length': '84', 'Connection': 'keep-alive', 'X-Frame-Options': 'SAMEORIGIN', 'Referrer-Policy': 'same-origin', 'Cache-Control': 'private, max-age=0, no-store, no-cache, must-revalidate, post-check=0, pre-check=0', 'Expires': 'Thu, 01 Jan 1970 00:00:01 GMT', 'Server': 'cloudflare', 'CF-RAY': '7e14e74a7df613f8-ORD', 'alt-svc': 'h3=":443"; ma=86400'\} \\
  
\hline
\hline
\end{tabular}
\end{table*}
According to the introduction on the OpenAI website and forum posts, there is no formal way to solve these running errors, so we wrote code to catch exceptions to ensure that the operation was not interrupted when an error occurred while using the API. 

The main code is as follows:

\begin{minipage}[l]{0.95\textwidth}
\lstset{texcl=true,basicstyle=\small\sf,commentstyle=\small\rm,mathescape=true,escapeinside={(*}{*)}}
\begin{lstlisting}
current_index = 0
while current_index<len(content_list):
    content = content_list[current_index]   
    try:
        response = openai.ChatCompletion.create(
            model="gpt-3.5-turbo",
            messages=[{"role": "user", "content": content}])
        answer = response['choices'][0]['message']['content']
        answers.append(answer)
        writer.writerow([paper_list[current_index],answer])
        current_index +=1
    except Exception as e:
        print(e)
        print("Sleeping......")
        time.sleep(60)
\end{lstlisting}
\end{minipage}

\subsection{GPT model outputs}
\label{sec_GPToutputs}
The GPT outputs in table~\ref{tab_ExampleGPT} show two examples that can aid to better understand the GPT model.
GPT provided comprehensive answers and an explanation for each assigned goal based on the prompts we designed.
As we set the max\_tokens parameter to 600, we were able to get the complete answers; if the parameter is set to a smaller value, such as 150, GPT will automatically truncate the answer regardless of the integrity of the answer.
Furethermore, we find that the output format is not uniform, as shown in table~\ref{tab_ExampleGPT}; the first output provides a confidence score and high, medium, and low information, while the second publication only gives the confidence score.
For the same prompt, the first output uses ``Goal," and the second output uses ``SDG."
By investigating more outputs, we noticed that for some publications, GPT failed to generate tags because of lack of sufficient information, while for some publications, GPT only gave high, medium, and low levels without specific confidence scores.
Some of the outputs generated by GPT also included goals with low-confidence values, such as, ``Goal 15: Life on Land - Although not explicitly mentioned in the publication, promoting renewable energy can indirectly contribute to the conservation of terrestrial ecosystems and biodiversity. Confidence level: Low (30\%)." In response, we conducted a comprehensive statistical analysis of the confidence values associated with all the results extracted from the publications. This analysis was undertaken to establish a discernible threshold that would enable the systematic removal of SDGs with low-confidence levels from the final set of results.

Based on the GPT outputs, categories were assigned as follows: Results with confidence levels less than 60\% (excluding) were classified as low confidence, those with confidence levels greater than or equal to 80\% were designated as high confidence, and outputs falling between 60\% and 80\% (excluding) were classified as having medium or moderate confidence.
Consequently, we established the threshold at 60\%, signifying that only goals exhibiting medium or higher confidence were retained as the conclusive output of the GPT model.
For later comparison and analysis, we also deleted the samples where the GPT model did not generate an answer or the answer only had the goal but no specific confidence score.
After processing, the output of 6,467 publications was finally retained for further comparative analysis.
We further processed the output to keep only the data information for comparison, so the GPT output of each publication only includes the target and specific confidence value, which is consistent with the output of the similarity measure, and the GPT model tagged 20,110 SDGs for 6,437 publications.

\begin{figure}[h]
\centering
\includegraphics[width=0.8\linewidth]{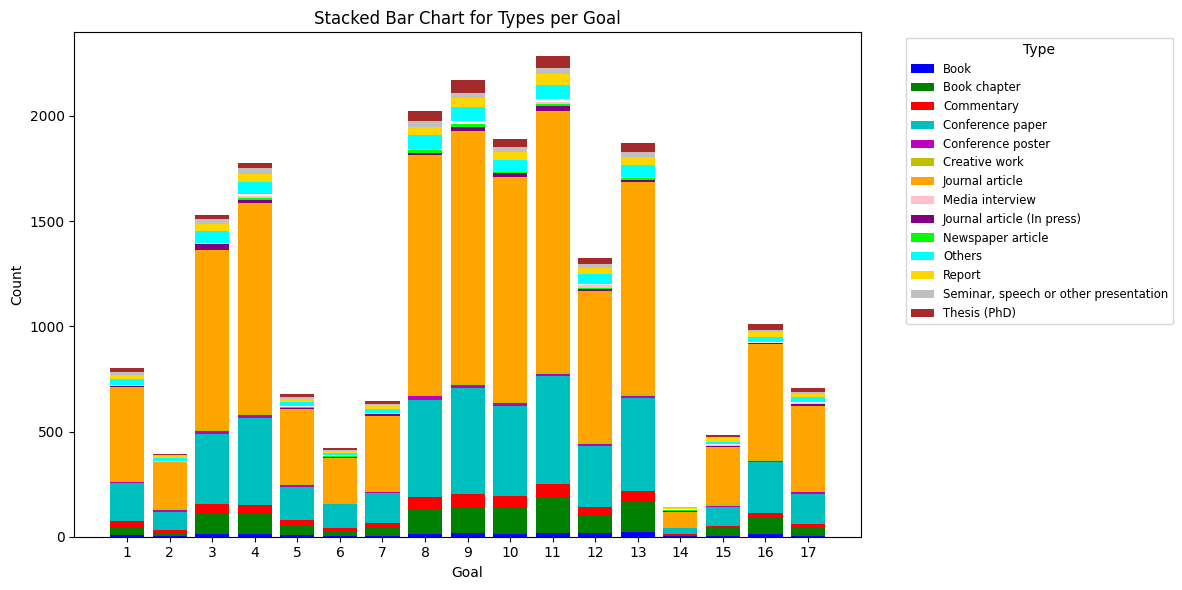}
\caption{\label{fig:GPT_barchart}Publication distribution across 17 SDGs: stacked bar chart of different publication types using GPT model.}
\end{figure}

\begin{table*}
\centering
\caption{Two example outputs for mapping publications to SDGs using the GPT API.}
\label{tab_ExampleGPT}

\begin{tabular}{m{1.5cm}|m{4.2cm}|m{9cm}}
\hline
\hline
Authors & Publication & Outputs\\ \hline

 Suni Mydock III et al. & Title: Influence of made with renewable energy appeal on consumer behaviour Abstract: The purpose of this paper is to explore the extent to which consumer purchasing behaviour is influenced by advertised information that a product is made with renewable energy .... ... in an Australian context, providing findings that assist both firms' and policy-makers' decision-making. & Based on the information provided, the publication aligns with the following Sustainable Development Goals (SDGs):1. \textbf{Goal 7}: Affordable and Clean Energy - confidence level: High (\textbf{100\%}) - The publication directly relates to the use and promotion of renewable energy.2. \textbf{Goal 12}: Responsible Consumption and Production - confidence level: High (\textbf{90\%}) - The publication addresses responsible consumption and production by highlighting the positive impact of renewable energy and sustainable business practices.3. \textbf{Goal 13}: Climate Action - confidence level: Medium (\textbf{70\%}) - Although the focus is on consumer behavior rather than direct climate action, the publication indirectly supports efforts towards mitigating climate change through promoting renewable energy.4. \textbf{Goal 9}: Industry, Innovation, and Infrastructure -confidence level: Medium (\textbf{60\%}) - While it does not directly address innovation and infrastructure, the publication suggests that firms can leverage renewable energy as a marketing strategy, which aligns with the goal of fostering innovation in industries.5. \textbf{Goal 15}: Life on Land - confidence level: Low (\textbf{30\%}) \\ \hline
Newton, Peter W. & Title: No sustainable population without sustainable consumption Abstract: In recent weeks, two major federal government strategy papers have been released: 'Our Cities, Our Future: A National Urban Policy for a Productive, Sustainable and Liveable Future' ... ... In a growing and urbanising 21st century world it is an unsustainable and inequitable nexus. & The publication aligns with the following SDGs:1. Sustainable Cities and Communities (\textbf{SDG 11}) - \textbf{90\%} confidence level: The publication discusses the need for sustainable urban development and the importance of accommodating population growth in cities. It also mentions the international recognition of Australia's cities' liveability.2. Responsible Consumption and Production (\textbf{SDG 12}) - \textbf{80\%} confidence level: The publication highlights the unsustainable and inequitable levels of resource consumption in Australia's cities. It emphasizes the need for sustainable consumption practices to achieve a sustainable population.The goals are listed in descending order of confidence level, with SDG 11 listed first. \\ 

\hline
\hline
\end{tabular}
\end{table*}
We performed statistics on the GPT outputs to see what trends GPT found in these publications.
Figure~\ref{fig:GPT_barchart} shows that these publications are most relevant to goals 11, 9, and 8; while the similarity measure finds goals 11, 12, and 4 most relevant.
\begin{itemize}
    \item Goal 11: Sustainable Cities and Communities
    \item Goal 9: Industry, Innovation and Infrastructure
    \item Goal 8: Decent work and economic growth
\end{itemize}

From figure~\ref{fig:BERT_barchart} and figure~\ref{fig:GPT_barchart}, we can confirm that both approaches map the most publications to Goal 11, Sustainable cities and communities, which is consistent with a research focus of Swinburne University of Technology.
The two methods did not agree on the second and third goals, so further processing is required.

\section{Comparison of two methods}
\label{sec:ComparisionTwoMethods}
So far, we have employed two methods for tagging publications to SDGs: similarity measure and the GPT model. Table ~\ref{tab_Cosine_ChatGPT} presents example outputs from both methods. According to the table, the outputs of the two methods vary, with instances of consensus and difference. Determining which output is more reasonable requires further exploration.
Considering that both are large-scale language models, the performance depends on the size of the training data set and the number of parameters used. GPT has 17.5 billion parameters, while BERT has 345 million parameters. GPT is pre-trained on massive text data and learns from a wide range of sources, including books, articles, websites, and other text resources on the Internet. 
Referring to the above features of the two models, training dataset size, and large-scale parameters, it is reasonable that the performance of GPT will exceed the similarity measure, so we used the output of GPT as a baseline to measure the performance of the similarity measure.

\begin{table*}
\centering
\caption{Comparison of SDG tagging results: similarity measure and GPT model.}
\label{tab_Cosine_ChatGPT}

\begin{tabular}{m{6cm}|m{4.7cm}|m{4cm}}
\hline
\hline
 Title & Similarity measure  & GPT \\ \hline

Linking scholarly articles to SDGs: an automated approach & {Goal 17: 0.39, Goal 4: 0.39, Goal 12: 0.38, Goal 15: 0.37, Goal 9: 0.36} & {Goal 4: 0.95, Goal 9: 0.85, Goal 16: 0.7, Goal 1: 0.6} \\ \hline
Dissemination of public health research to prevent non-communicable diseases: a scoping review & {Goal 3: 0.38}&{Goal 3: 0.9} \\ \hline 
 Transitioning cities and post-COVID planning & {Goal 11: 0.60, Goal 13: 0.46, Goal 15: 0.40, Goal 3: 0.40, Goal 6: 0.38} & {Goal 11: 0.80, Goal 13: 0.70, Goal 9: 0.60} \\ \hline
Everyday humanitarianism during the 2019/2020 Australian bushfire crisis & {Goal 11: 0.44, Goal 13: 0.38, Goal 15: 0.37}&{Goal 11: 0.9, Goal 13: 0.85, Goal 1: 0.7, Goal 3: 0.65, Goal 15: 0.6} \\ \hline 
Data for Good Collaboration: research report & {Goal 11: 0.39, Goal 4: 0.37, Goal 1: 0.36} &{Goal 4: 0.85, Goal 17: 0.8, Goal 9: 0.75, Goal 10: 0.7, Goal 3: 0.65, Goal 8: 0.6} \\ \hline 
\hline
\end{tabular}
\end{table*}

\subsection{Evaluation metrics}
A confusion matrix is a matrix that breaks down correctly and incorrectly classified information into:
\begin{itemize}
    \item True positive (TP): Correctly predicting the positive class
    \item True Negative (TN): Correctly predicting the negative class
    \item False Positive (FP): Incorrectly predicting the positive class
    \item False Negative (FN): Incorrectly predicting the negative class
\end{itemize}

\begin{figure}[h]
\centering
\includegraphics[width=0.6\linewidth]{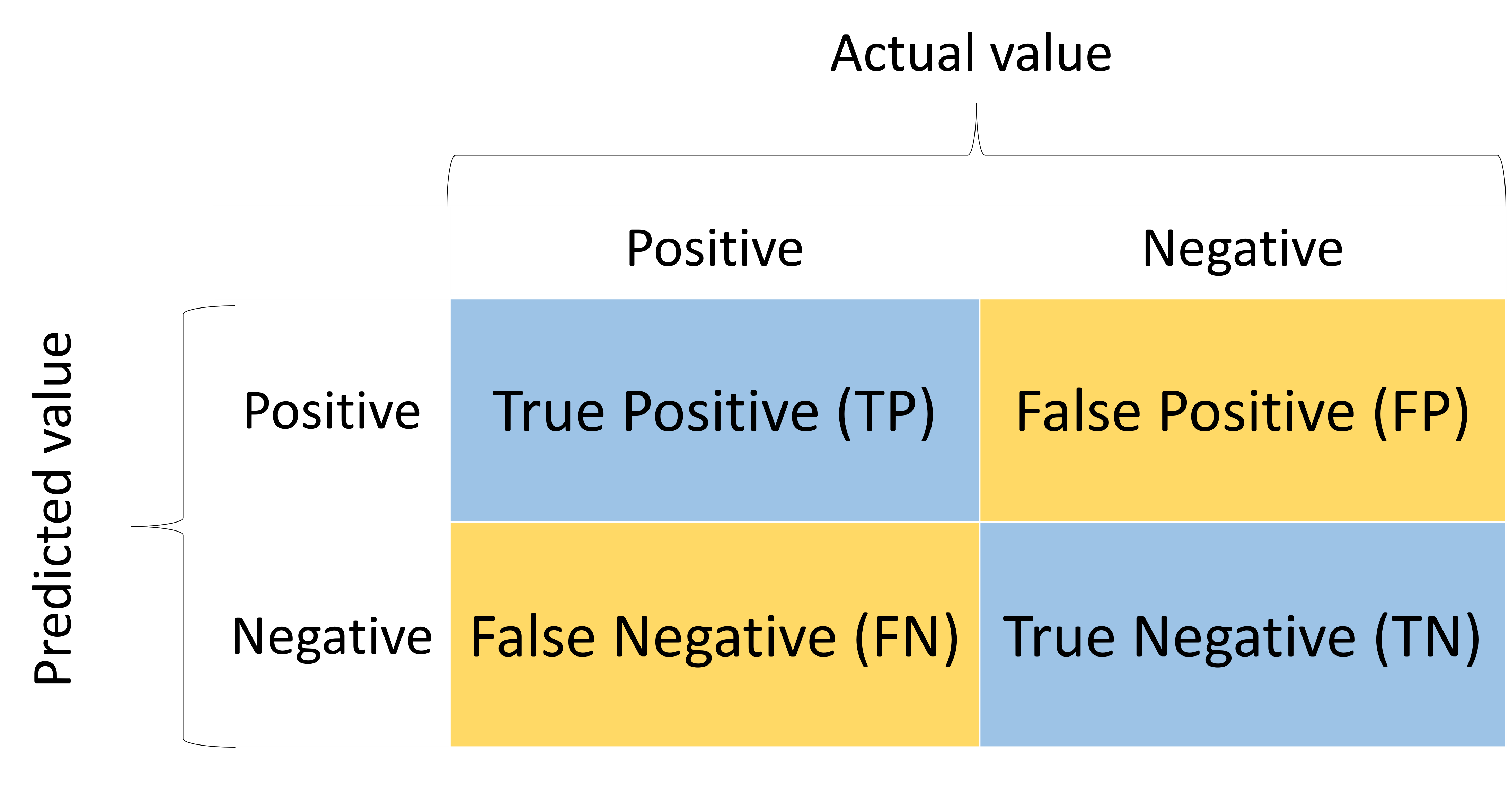}
\caption{\label{fig:Confusion}Confusion Matrix.}
\end{figure}

SDG mapping is a multi-label problem and the accuracy cannot be relied on to determine the model's performance; that is, the labels predicted by the model for each record need to be entirely consistent with the actual labels. 
What needs to be calculated is the intersection of the predicted labels and the actual labels; that is, the predicted labels are a subset of the actual labels, or they also contain some wrong labels, or there is no intersection (the predicted labels are all wrong).
Therefore, the precision, recall, and F1 scores listed below are more suitable for evaluating model performance.

\begin{equation}
\label{prf}
\begin{array}{c}
Precision= {TP}/{(TP+FP)}\\
\\
Recall= {TP}/{(TP+FN)}\\
\\
F_{1}=2*\frac{(Precision \times Recall)}{Precision + Recall}.\\
\end{array}
\end{equation}

We evaluated the multi-label classification models by generating a confusion matrix for each class label using sklearn's multilabel\_confusion\_matrix function.
Aggregate metrics including macro, micro, weighted, and sampled avg give a high-level view of how our model performed over all classes.
The classification report is shown in table~\ref{tab:ClassificationReport}.

\begin{table}[]
\centering 
\caption{Performance evaluation of multi-Label classification model: classification report.}
\label{tab:ClassificationReport}
\begin{tabular}{m{2.4cm}|m{2cm}|m{2cm}|m{2cm}|m{2cm}}

\rowcolor[HTML]{BDD7EE} 
{\color[HTML]{BDD7EE} \textbf{..}} & \textbf{Precision} & \textbf{Recall} & \textbf{F1-score} & \textbf{Support} \\
\rowcolor[HTML]{EDEDED} 
SDG 1                              & 0.386642           & 0.516981        & 0.442411          & 795              \\
SDG   2                            & 0.439481           & 0.778061        & 0.561694          & 392              \\
\rowcolor[HTML]{EDEDED} 
SDG 3                              & 0.743265           & 0.305737        & 0.433256          & 1534             \\
SDG   4                            & 0.670011           & 0.718573        & 0.693443          & 1766             \\
\rowcolor[HTML]{EDEDED} 
SDG 5                              & 0.641548           & 0.463235        & 0.538002          & 680              \\
SDG   6                            & 0.339265           & 0.682578        & 0.453249          & 419              \\
\rowcolor[HTML]{EDEDED} 
SDG 7                              & 0.527886           & 0.636933        & 0.577305          & 639              \\
SDG   8                            & 0.573699           & 0.392292        & 0.465962          & 2024             \\
\rowcolor[HTML]{EDEDED} 
SDG 9                              & 0.62665            & 0.460083        & 0.530601          & 2167             \\
SDG   10                           & 0.484058           & 0.17616         & 0.258314          & 1896             \\
\rowcolor[HTML]{EDEDED} 
SDG 11                             & 0.689772           & 0.717678        & 0.703448          & 2274             \\
SDG   12                           & 0.512136           & 0.802281        & 0.625185          & 1315             \\
\rowcolor[HTML]{EDEDED} 
SDG 13                             & 0.658367           & 0.352721        & 0.459347          & 1874             \\
SDG   14                           & 0.15512            & 0.741007        & 0.256538          & 139              \\
\rowcolor[HTML]{EDEDED} 
SDG 15                             & 0.337411           & 0.69392         & 0.454047          & 477              \\
SDG   16                           & 0.481994           & 0.343874        & 0.401384          & 1012             \\
\rowcolor[HTML]{EDEDED} 
SDG 17                             & 0.20093            & 0.305516        & 0.242424          & 707              \\
micro   avg                        & 0.524889           & 0.493933        & 0.508941          & 20110            \\
\rowcolor[HTML]{EDEDED} 
macro avg                          & 0.498131           & 0.534567        & 0.476271          & 20110            \\
weighted   avg                     & 0.56906            & 0.493933        & 0.501287          & 20110            \\
\rowcolor[HTML]{EDEDED} 
samples avg                        & 0.732922           & 0.524073        & 0.52673           & 20110           
\end{tabular}
\end{table}

\subsection{Classification report analysis}
The multi-label classification task is challenging in the NLP field. 
There are a total of 17 categories in this study, increasing the classification difficulty. 
According to the report, the performance of the proposed method on the 17 goals varies widely, and we summarize the common scenarios as follows.
\begin{itemize}
    \item Higher precision values \& lower recall values: The model is conservative in making positive predictions. It is cautious and only predicts positive labels when relatively confident about them. It does not predict every expected label but also does not predict extra labels, such as goals 3, 5, 8, 9, 13.
    \item High precision values \& high recall values:  This is an ideal scenario where the model is accurate and comprehensive in its predictions. The model can predict positive labels with high accuracy and captures a large portion of actual positive instances, such as goals 4, 11.
    \item Low precision values \& high recall values: The model may predict positive labels for many instances, but it is also likely to produce a significant number of false positives, leading to a drop in overall accuracy, such as goals 1, 2, 6, 12, 14, 15.
    \item Low Precision \& low Recall: The model struggles to predict positive labels (low recall) accurately and makes numerous incorrect positive predictions (low precision), resulting in overall poor performance, such as goals 7, 10, 16, 17.
\end{itemize}

The proposed method performs best on goals 4 and 11 with high precision and recall but performs poorly on other objectives. 
The four averaging metrics in multi-label classification reports offer different perspectives on model performance. The macro average calculates metrics for each label and averages them equally, giving a balanced view. The micro average aggregates performance across all instances, providing an overall performance measure. Weighted average adjusts for class imbalances by considering class support when averaging. Sampled average treats instances as labels, providing insights into individual instance performance. The choice of averaging method depends on whether you seek a class-balanced overview, an instance-level perspective, or adjustments for class imbalances in the evaluation. 
Considering the unbalanced number of samples in each goal, we only consider the performance of a single goal, not the average.

\section{Hybrid approach to get a high confidence result}
\begin{figure}[h]
\centering
\includegraphics[width=0.6\linewidth]{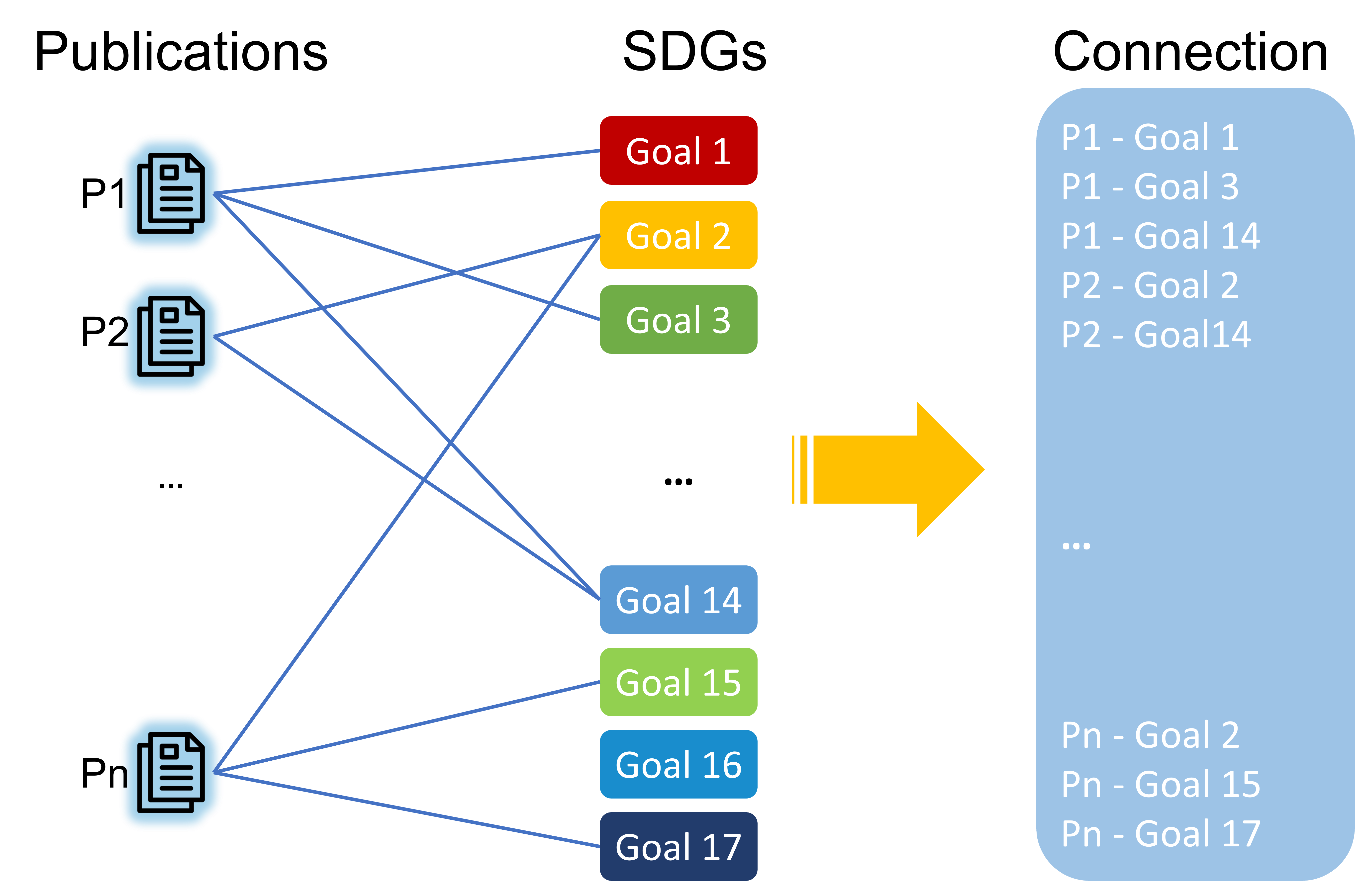}
\caption{\label{fig:Connection}The diagram shows how the connections are constructed between publication and SDGs.}
\end{figure}

There are differences between the conclusions drawn from the similarity measure and the GPT model about mapping Swinburne University of Technology publications to the SDGs. 
In the following section, we employed set theory to explore the intersection of these two results, an approach that will yield high-confidence, well-established results.
Formally, we describe this process as follows: given two distinct sets, denoted set A and set B, set A corresponds to the results obtained from the similarity measure, 18,924 SDGs for 7,403 publications. While set B contains the results from the GPT model, that is, 20,110 SDGs for 6,467 publications. 
The elements of each set are connections between publications and assigned SDGs, figure~\ref{fig:Connection} shows the process of establishing connections between publications and SDGs.
The intersection of sets A and B can be formally expressed using set notation:

\begin{equation}
  A \cap B = \left\{connection: connection \in A \text{ and } connection \in B\right\}
\end{equation}

\begin{figure}[h]
\centering
\includegraphics[width=0.3\linewidth]{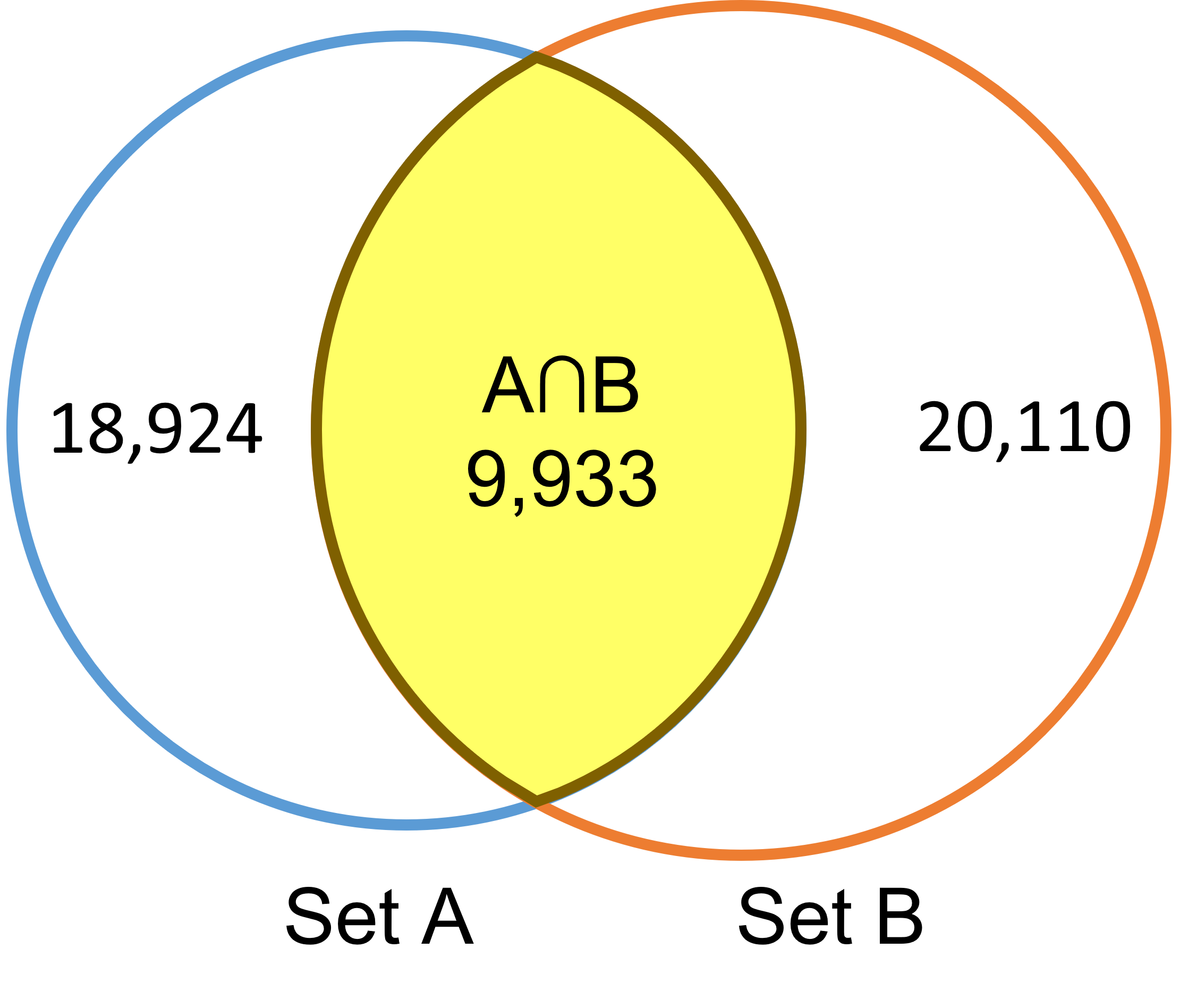}
\caption{\label{fig:Overlapping} Intersection of sets using a Venn diagram. Set A corresponds to the outputs obtained by the similarity measure, while set B contains the outputs from the GPT model, the intersection is 9,933 connections.}
\end{figure}

\begin{figure}[h]
\centering
\includegraphics[width=1\linewidth]{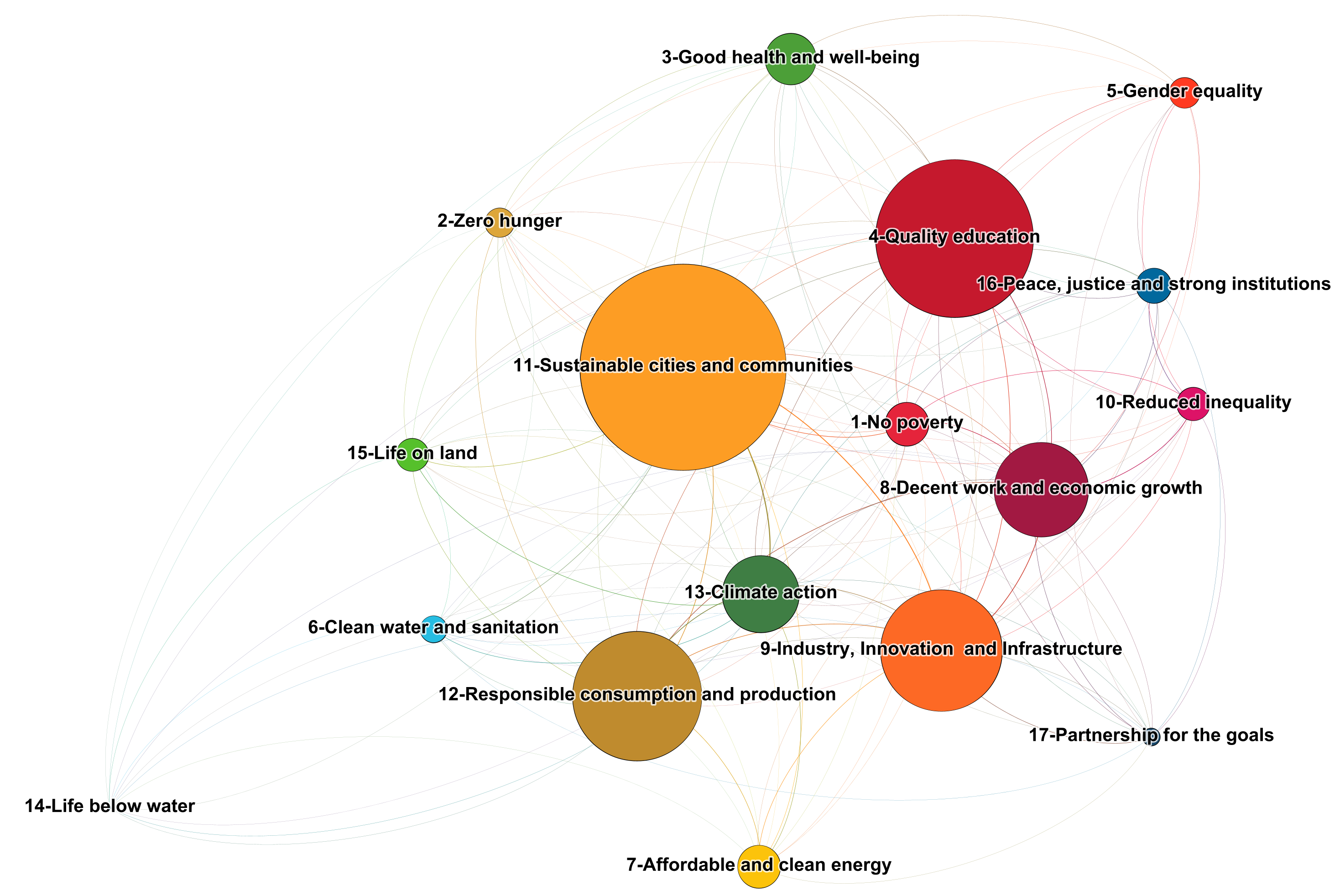}
\caption{\label{fig:Swin_graph}Statistics of Swinburne publications in 17 SDGs using the hybrid method. The size of the node corresponds to the number of publications. The thicker the connections between nodes, the more often the two goals appear in one publication.}
\end{figure}

Figure~\ref{fig:Overlapping} shows the Venn diagram of intersection, with a total of 39,034 connections, of which 9,933 connections are the same, accounting for 34.13\% of the entire connections.
These duplicate connections belonged to 6,136 publications, meaning that the two methods reached a consensus on SDG labels for 6,136 publications (with at least one label), accounting for 82.89\% of all publications.
The hybrid of the two methods ensures the accuracy of publication tags, and we further counted the distribution of these publications in 17 SDGs to gain an insight into Swinburne University of Technology's current status in achieving the SDGs, as shown in figure~\ref{fig:Swin_graph}.

\label{sec:hibrid}

The hybrid approach shows that most publications are assigned to goal 11, Sustainable cities and communities, and goal 4, Quality education.

Swinburne University of Technology has invested substantial efforts in SDG 11, focusing on creating inclusive, safe, resilient, and sustainable cities and human settlements. The Swinburne Innovative Planet Research Institute applies innovative and socio-technical approaches to address the challenges facing the world's fast-growing cities - challenges such as climate change, altered energy, food or water security, issues of mobility and health, aging population and disruptions to the nature of work and home. Moreover, SDG 4 is dedicated to ensuring inclusive and equitable quality education, fostering lifelong learning opportunities, and advancing knowledge dissemination for sustainable development. 
In contrast, the number of publications for SDG 14, ``Life below water", was low.

\section{Discussion}
%Amir: Discussion should address what this paper is achieved and what are the future work.
In this study, we applied Artificial Intelligence technology to map publications to SDGs. Mapping of this type is designed to support governments, businesses, and research institutions to identify their past and current research related to SDGs and to inform their research strengths and impact.
We collected 82,649 publications from the Swinburne University of Technology open-access repository and performed an initial screening of these publications using similarity measures (AI-based technique). The initial screening identified 9\% of publications (7,403) as potential candidates for SDG-related work.
Using the similarity score, we assigned the papers to SDGs. In the next phase, we leveraged the GPT model to remap these publications to SDGs, aiming for an alternative approach to mapping and gaining better insights into the performance of different models. 
The utilization of the GPT model assigned 6,437 publications to various SDGs but failed to provide answers for 966 publications.
One advantage of using GPT (and potentially other LLM models) is providing insights into the performance of the more traditional classification methods such as (similarity measures).
% Our case study results show that the similarity measure performs well when classifying publications into goal 11 and goal 4, with high precision and recall values. However, this is a limited case study, and the results might be affected by environmental factors such as the university's research priorities. 

In pursuing potentially more reliable results, we adopted a hybrid strategy that amalgamated the outputs of both approaches to effectively identify relevant Sustainable Development Goals (SDGs) for publications.
Both approaches reached a consensus on SDGs for 6,136 publications (with at least one tag), which accounted for 82.89\% of pre-selected publications. This analysis proved that the similarity measure achieves comparable performance to the GPT model.
We further perform a statistical analysis on the 6,136 publications, revealing that Swinburne University prioritizes its efforts towards SDG 11, which pertains to Sustainable Cities and Communities and SDG 4, focused on Quality Education; these findings are consistent with the classification experiment. In contrast, SDG 14, which addresses Life below Water, received relatively little attention.
Our research demonstrates that an AI-based similarity measurement can attain performance comparable to GPT's in mapping publications to Sustainable Development Goals (SDGs).
Considering that the GPT API is not easy to implement for non-IT researchers, and there are confidential documents that cannot be made exposed to GPT API (e.g., government projects and reports), the similarity measures we propose provide a good alternative.

A subsequent phase of the research roadmap involves training a supervised machine learning model to classify multi-label publications. The model will utilize the results of the hybrid method as ground truth labels. Subsequently, we intend to deploy the model in other research institutions for broader applications.
Further investigation is also being contemplated, considering that each goal encompasses two specific components: targets and indicators (17 goals, 169 targets, and 231 distinct indicators). We aim to extend our predictions to finer-grained categories, aiming to classify publications into the corresponding targets and indicators within each goal.

% \section*{CRediT authorship contribution statement}
% \textbf{Hui Yin}: Conceptualization, Methodology, Experiment, Writing - Original Draft. 
% \textbf{Xiao Liu}: Project administration, Supervision, Writing - Review \& Editing. 
% % \textbf{Xiao Liu}: Supervision, Writing - Review \& Editing.
% \textbf{Yutao Wu}: Prepare figures and tables. \textbf{Hilya Mudrika Arini}: Writing - Review \& Editing. 
% \textbf{Rami Mohawesh}: Review \& Editing. 
% % % All authors reviewed the manuscript.
\section*{Acknowledgement}
This research was supported by the Australian Government through the Australian Research Council's Industrial Transformation Training Centre for Information Resilience (CIRES) project number IC200100022.

\section*{Declarations}
\textbf{Conflict of interest} The authors declare that they have no known competing financial interests or personal relationships that could have appeared to influence the work reported in this paper.

\textbf{Data availability}
The data that support the findings of this study are available upon reasonable request.

\bibliography{sn-bibliography}% common bib file

\end{document}